\documentclass[letterpaper, 10 pt, conference]{ieeeconf}
\IEEEoverridecommandlockouts                              

\usepackage{float}
\usepackage{dblfloatfix}
\usepackage{amssymb}
\usepackage{amsmath}
\usepackage{lipsum}
\usepackage{algorithm}
\usepackage[noend]{algorithmic}

\usepackage{graphicx}
\usepackage{subcaption}
\usepackage{todonotes}

\title{\LARGE \bf
A Conceptual Framework for AI-based Decision Systems in Critical Infrastructures}

\author{Milad Leyli-abadi$^{1}$, Ricardo J. Bessa$^{2}$, Jan Viebahn$^{3}$, Daniel Boos$^{4}$, Clark Borst$^{5}$, Alberto Castagna$^{6}$\\ Ricardo Chavarriaga$^{7}$, Mohamed Hassouna$^{12}$, Bruno Lemetayer$^{8}$, Giulia Leto$^{5}$, Antoine Marot$^{8}$, Maroua Meddeb$^{1}$\\Manuel Meyer$^{9}$, Viola Schiaffonati$^{10}$, Manuel Schneider$^{9}$, Toni Waefler$^{11}$
\thanks{*The research leading to this work is part of the AI4REALNET (\textit{AI for REAL-world NETwork operation}) project, which received funding from European Union’s Horizon Europe Research and Innovation programme under the Grant Agreement No 101119527, and from the Swiss State Secretariat for Education, Research and Innovation (SERI). This project is funded by the European Union and SERI. Views and opinions expressed are however those of the author(s) only and do not necessarily reflect those of the European Union and SERI. Neither the European Union nor the granting authority can be held responsible for them.}
\thanks{$^{1}$IRT SystemX, $^{2}$INESC TEC, $^{3}$TenneT, $^{4}$SBB, $^{5}$Delft University of Technology, $^{6}$EnliteAI, $^{7}$Zurich University of Applied Science, $^{8}$Réseau de Transport d'Electricité, $^{9}$Flatland Association, $^{10}$Politecnico di Milano, $^{11}$University of Applied Sciences Northwestern Switzerland, $^{12}$Fraunhofer IEE
        }%
}

\begin{document}

\maketitle
\thispagestyle{empty}
\pagestyle{empty}

\begin{abstract}

The interaction between humans and AI in safety-critical systems presents a unique set of challenges that remain partially addressed by existing frameworks. These challenges stem from the complex interplay of requirements for transparency, trust, and explainability, coupled with the necessity for robust and safe decision-making. A framework that holistically integrates human and AI capabilities while addressing these concerns is notably required, bridging the critical gaps in designing, deploying, and maintaining safe and effective systems. This paper proposes a holistic conceptual framework for critical infrastructures by adopting an interdisciplinary approach. It integrates traditionally distinct fields such as mathematics, decision theory, computer science, philosophy, psychology, and cognitive engineering and draws on specialized engineering domains, particularly energy, mobility, and aeronautics. Its flexibility is further demonstrated through a case study on power grid management.

\end{abstract}

\section{Introduction}
\label{sec:introduction}

Artificial Intelligence (AI) is showing high potential to transform the management of critical infrastructures~\cite{McMillan2022}, tackling pressing challenges like climate change and the rising demand for energy and mobility systems while advancing strategic objectives such as energy transition and digital transformation. On the other hand, integrating AI in critical sectors introduces significant challenges, many of which are already being addressed by emerging regulatory frameworks, such as the European Union AI Act. These frameworks emphasize the importance of safety, transparency, and adherence to ethical standards and principles to mitigate a wide range of risks, including technical, social, and environmental hazards associated with deploying AI in high-risk domains. Another key challenge lies in fostering effective human-AI collaboration. This involves integrating human expertise into AI-based systems and promoting adaptive learning and collaborative decision-making processes. Various approaches can be explored to achieve this, such as co-learning frameworks where both AI and humans evolve their knowledge together or systems in which AI learns independently but is designed to work alongside humans in joint decision-making. 

In response to these challenges, this paper proposes a conceptual framework for AI-based systems in critical infrastructures. This framework integrates emerging AI algorithms, open-source digital environments optimized for AI, and socio-technical design principles and practices for responsible research and innovation that support human-machine interaction. It aims to improve real-time and predictive operations across key sectors such as power networks, railways, and air traffic management -- identified as vital to Europe and prioritized in national AI strategies worldwide.

Existing decision-making frameworks provide valuable insights into AI integration but often operate in isolation, focusing on specific sectors or challenges (see Section~\ref{sec:related_works} for a review). There is a need for a unified approach that balances technical, ethical, and human considerations in critical infrastructures. This paper proposes a novel framework that fosters collaboration between human control and AI, ensuring AI enhances rather than replaces human decision-making.

Designed with real-world, safety-critical applications in mind, the proposed framework prioritizes transparency, trust, and accountability, improving both social and technical performance. By directly addressing key challenges such as reliability, security, and response time in complex environments, it supports adaptable, ethically aligned, and context-aware human-AI systems for safer and more effective decision-making.


A review of existing frameworks is provided in Section \ref{sec:related_works}. The proposed conceptual framework is described in Section \ref{sec:conceptual_framework}. An example of its instantiation is provided in Section \ref{sec:instantiation}, and Section \ref{sec:conclusions} concludes the paper. 


\section{Related Work}
\label{sec:related_works}
The decision-making processes within critical infrastructures are increasingly supported by frameworks that integrate AI to enhance reliability, efficiency, and safety. These frameworks aim to address the unique challenges of critical sectors 
by providing structured approaches to managing complexity and uncertainty. Herein, we review some of the prominent frameworks developed for decision-making in the energy and mobility domains.

\paragraph{Decision support systems in power grids}
Power grid evolution driven by decarbonization increases operational complexity, requiring enhanced control room supervision. 
When managing technical problems such as congested lines, power system engineers rely on specialized expertise, real-time and forecasted data, and simulation tools for complex decision-making. Yet, they have limited access to automated decision-support tools, requiring manual exploration and experience-driven simulations to implement grid adjustments and remedial actions.

Marot et al. \cite{Marot2022} propose an AI-based agent framework to assist power grid operators by ensuring alarms are sent in advance when action confidence is low, preventing human-out-of-the-loop scenarios. It emphasizes \textit{credibility} through transparency, \textit{reliability} via consistent performance, and \textit{intimacy} by explaining incorrect actions to build trust. Greitzer et al. \cite{Greitzer2008} introduce a naturalistic decision-making framework that models interactions between real-world cues and operator cognition using mental simulations to anticipate control actions.

Fan et al. \cite{Fan2024} propose a human-machine hybrid intelligence framework for power grid dispatching, integrating AI, a digital twin, and human operators to enable bidirectional learning and dynamic task reassignment. Hilliard et al. \cite{Hilliard2024} develop the Work Domain Analysis framework, linking grid assets to high-level objectives through means-ends relationships and cause-effect models. Finally, GridOptions~\cite{Viebahn24} is the first AI-based decision-support tool deployed in a Transmission System Operator (TSO) control room, applying Evaluative AI~\cite{10.1145/3593013.3594001} to enhance decision-making through quality-diversity multi-objective optimization. That is, by providing evidence for and against a range of possible options (instead of providing recommendations that can only be accepted or rejected), it leverages human expertise in decision-making and mitigates issues of over and under-reliance.

\paragraph{Railway traffic management systems}
The traffic management in modern railway networks is becoming ever more complex due to dense mixed traffic and increasing passenger and freight demand. Rescheduling and rerouting trains present a significant challenge when deviations from the planned schedule occur, such as delays or infrastructure disruptions. While these tasks are traditionally performed by human operators, they receive support from traffic management systems (TMS) with various degrees of automation, like the Swiss Federal Railway's (SBB) Rail Control System (RCS)~\cite{sbb_rail_2020} which provides real-time traffic monitoring and dynamic routing options.

While traditional TMS in the railway sector rely heavily on operations research, there is growing interest in leveraging machine learning, particularly multi-agent reinforcement learning (MARL), for traffic management. Efforts in this area range from enhancing solution quality and response times for dynamic routing---such as in German Railway’s (DB) automatic dispatching assistant
---to exploring novel rescheduling and rerouting approaches using MARL~\cite{stefan_schneider_intelligent_2024} within simulation environments like Flatland~\cite{laurent_flatland_2021}. Additionally, recent public and private open-source research initiatives in the railway sector continue to drive innovation in TMS~\cite{sncf_open_2022,ora_2024,europes_rail_fp1-motional_2023}.

\paragraph{Air Traffic Management (ATM) systems}
ATM consists of several entities that all need to work together seamlessly to achieve safe and efficient air traffic operations. Those entities operate at different time scales, ranging from long-term strategic flight planning (years to months prior to operation), pre-tactical operations (days to hours before operation) towards tactical operations in the execution phase of flight. Currently, the ATM system is human-centric, meaning that human operators (ranging from strategic flight planners toward tactical air traffic controllers (ATCOs) bear the ultimate responsibility for operational safety.

The steady growth of air traffic, projected at 2\% annually, has exposed a gap between future demand and the capacity of current ATM infrastructure \cite{EUROCONTROL2024e}. To address this, the ATM community is increasingly adopting automated support tools. While airspace capacity is geometrically linked to minimum safety separation distances \cite{majumdar2002estimation}, its primary constraint is the ability of ATCOs to monitor airspace on the day of operations (O-Day) \cite{majumdar2002estimation}. Since airspace capacity depends on the sustainable workload of ATCOs, the key challenge lies in scaling their availability to match traffic growth, making training and certification efforts crucial.

Beyond decision-support tools aimed at reducing ATCO workload (e.g., \cite{zohrevandi2022design,westin2020building,nunes2021human}), another approach focuses on Air Traffic Flow and Capacity Management (ATFCM) at the strategic level, up to seven days before O-Day. This allows human operators ample time to review automated recommendations. ATFCM involves optimizing flight distribution and ATCO assignments to balance workload. Martin et al. \cite{martin2023stam} proposed using adjacent unused sectors to increase capacity, while other studies address strategic deconfliction by minimizing predicted conflicts through trajectory planning \cite{chaimatanan2013strategic}. Additionally, research explores dynamic airspace sector structures, using heuristic and learning-based methods to optimize workload distribution among controllers \cite{xue2009airspace,gerdes2018dynamic}.

\paragraph{Socio-technical systems and Human-AI interaction} Frameworks that address the socio-technical aspects of decision-making are gaining prominence. As an example, the Joint Control Framework focuses on the shared decision-making process between humans and AI-based systems \cite{lundberg2021framework}. It outlines strategies for co-learning and adaptive control, ensuring that human expertise is integrated into AI-driven decisions, particularly in critical scenarios where full automation may not be feasible.

\paragraph{Trustworthy AI frameworks} Trust is a cornerstone of decision-making frameworks in critical infrastructures. The Confiance.AI framework \cite{gelin2024confiance} and the Human AI Ethical Framework \cite{dignum2019humane} emphasize the development of AI-based systems that are not only effective but also transparent, ethical, and aligned with human values in order to be deemed trustworthy. These frameworks provide guidelines for ensuring that AI-based decision systems are trustworthy and capable of supporting decisions in critical environments.

\paragraph{Risk management and regulations} The Assessment List for Trustworthy Artificial Intelligence (ALTAI) and the regulatory provisions of the AI Act by the European Union offer frameworks for assessing and managing risks associated with AI-based systems in critical infrastructures~\cite{fedele2024altai}. These frameworks provide a structured approach for evaluating the ethical and functional aspects of AI, ensuring that decision-making processes meet stringent safety and accountability standards. Nonetheless, these frameworks focus mainly on assessment of safety and trustworthy aspects of AI-systems and further specialization for critical systems is still required. 

This work complements existing CPHS modeling approaches \cite{netto2017special}, including SysML-based architectures and agent-based models, by offering a cross-disciplinary conceptual foundation that incorporates cognitive and ethical dimensions often absent from formal system design frameworks.

\begin{figure}
    \centering
    \includegraphics[width=\columnwidth]{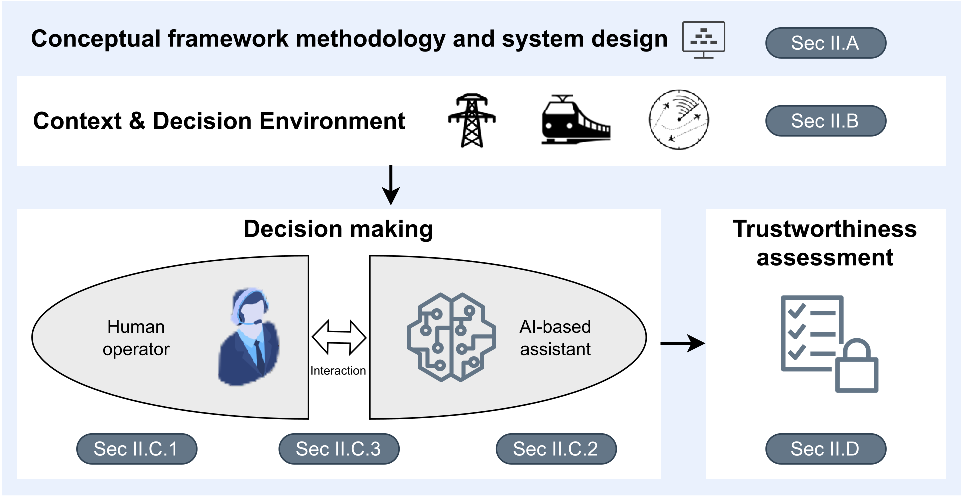}
    \caption{The proposed conceptual framework, comprising three main components governed by system design principles and their corresponding sections}
    \label{fig:ai4realnet_scheme}
\end{figure}

\section{Proposed Conceptual Framework}
\label{sec:conceptual_framework}
A high-level overview of the proposed conceptual framework for critical infrastructures is illustrated in Figure \ref{fig:ai4realnet_scheme}, which also outlines the structure of this section. We adopt an interdisciplinary approach to develop the conceptual framework for critical infrastructures by integrating different fields such as psychology, philosophy, ethics and cognitive engineering. The framework also drew on mathematics, decision theory, computer science, and specialized engineering domains, particularly energy and mobility. 

The systems engineering and theories are adapted for trustworthy AI integration in designing the conceptual framework's operational, functional, and logical architectures to meet both functional and non-functional requirements of critical infrastructures. 
Based on the observed context and decision environment, the decision-making process is responsible 
for ensuring the stability and resilience of the corresponding critical infrastructure. The decisions may result from collaboration between human operators and AI-based assistants through a provided interface. 
Furthermore, the decisions must comply with established trustworthy key performance indicators (KPIs) and undergo validation by a regulatory authority. The subsequent sections provide a detailed explanation of each component of the framework.

\subsection{Conceptual Framework Methodology and System Design}
Systems engineering principles and theories provide a structured approach to designing the proposed conceptual framework for decision-making in critical infrastructures by integrating interdisciplinary methodologies, risk management strategies, and regulatory compliance mechanisms. To achieve this,  we adopted a model-based systems engineering approach using the ARCADIA method \cite{roques2016mbse}, introducing the following analyses and views.

\begin{figure}[H]
    \centering
    \includegraphics[width=0.9\linewidth]{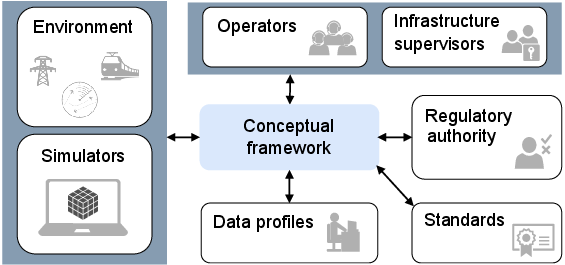}
    \caption{Operational environment diagram. External entities, interactions, and contexts in which a system operates}
    \label{fig:env_diag}
\end{figure}

\paragraph{Operational view} The operational view answers the question: ``\textit{What external elements interact with the system, and how do they influence it?}'' and focuses on 
how the system functions in real-world scenarios, emphasizing interactions with stakeholders such as operators, regulatory agents, and supervisors. The operational environment diagram in Figure \ref{fig:env_diag} visually represents these external entities, conditions, and interactions that influence the system. It helps define the system boundaries, showing how the system interacts with its environment, including users, other systems, hardware, software, and external constraints such as regulations or physical conditions. These stakeholders ensure the system's secure and efficient operation, aided by tools like simulators and AI-based assistants supported by data profiles that enhance decision-making and compliance. The framework is designed to handle various operational contexts, integrating human-in-the-loop mechanisms for collaborative decision-making between human operators and AI-based systems.

\begin{figure}
    \centering
    \includegraphics[width=1\linewidth]{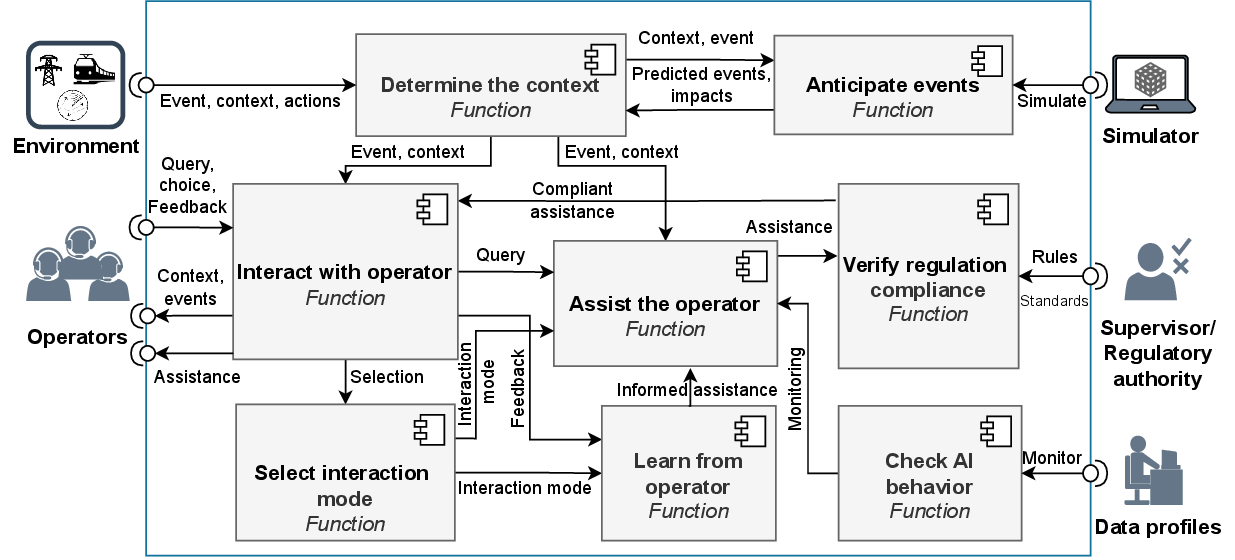}
    \caption{Functional interaction diagram. Eight identified core functions along with their interactions and information flow}
    \label{fig:func_inter_diag}
\end{figure}

\paragraph{Functional view} The functional view answers the question: ``\textit{How does the system achieve its objectives?}''. Thereby, it defines the system functions and shows how system components process inputs, produce outputs, and interact. The functional view outlines eight key functions, which are shown in Figure \ref{fig:func_inter_diag} along with their interactions and information flow. 
It involves determining real-time context by analyzing both internal (simulator) and external (environment) data, anticipating future events to proactively address potential issues, and interacting with the operator to facilitate communication and support. It also includes obtaining feedback from the operator on the provided assistance, selecting human-AI interaction modes to customize collaboration, learning from the operator's actions to enhance the system's knowledge base, assisting with decision-making, ensuring compliance with regulations to maintain trustworthiness, and continuously monitoring AI behavior to detect and respond to anomalies.

Identifying the different contextual and interaction scenarios ensures the system responds efficiently to contextual changes, supports informed decision-making, and upholds regulatory standards. The seamless flow of information between these functions and external entities promotes continuous learning and robust performance, ensuring the system remains resilient and effective in real-world scenarios.

\paragraph{Logical view} The logical view answers the question: ``\textit{How are the system’s functions structured as components and how do they interact logically?}'' and provides an abstract structure 
for integrating the identified core functions, focusing on seamless interaction among subsystems and stakeholders identified in the previous views. 
This view forms the foundation for designing the whole decision-making process, as illustrated in Figure \ref{fig:decision-making}. As depicted in this figure, the logical view of the conceptual framework provides a structure integrating five key components and their interactions. The decision environment provides context information to both human and AI-based agents, which process this data internally and enhance their capabilities through an interactive interface. The AI's behavior is continuously monitored and evaluated based on properties relevant to the decision-maker's perspective, ensuring it effectively supports the decision-making. To uphold trustworthiness in the final decision, a regulatory officer assesses compliance with established standards. A more detailed explanation of these diagrams is provided in Section \ref{sec:decision_making}.  

\subsection{Context and Decision Environment}
Decision-making in critical network infrastructures is a complex process influenced by external events, such as disruptions or emergencies, and the constraints of network capacity. These decisions involve multiple stakeholders across various time horizons and must balance trade-offs between conflicting objectives under tight time constraints. They are critical because these infrastructures underpin vital societal functions, including safety, health, and economic stability. The decision-making 
is triggered by detected events and involves iterative interactions between human operators and AI-based systems, blending manual and autonomous actions. Preventive and corrective measures are selected within a large action space, often in real-time, to ensure infrastructure resilience.


Analysis across three domains---railways, air traffic, and electricity---reveals that while decision contexts vary due to domain-specific factors, there is significant similarity in decision characteristics 
and evaluation criteria, such as trust in AI-based systems and assistant relevance. However, impacts remain largely domain-specific, reflecting unique operational priorities. Multi-domain studies highlight shared methodologies and potential for collaboration, with the strongest similarity observed between railway and air traffic domains. These insights emphasize the importance of cross-domain approaches in enhancing the effectiveness and robustness of decision-making processes in critical network infrastructures.

\subsection{Decision-making Process}
\label{sec:decision_making}
The decision-making process in critical systems involves a dynamic interplay between human expertise, AI-based decision-making capabilities, and their collaborative interaction. This section explores three key dimensions: human decision-making, which leverages domain knowledge and intuition; AI-based decision-making, which provides data-driven insights and scalability; and human-AI interaction, which integrates these strengths to optimize decisions under complex and uncertain conditions.

\begin{figure*}[htbp]
\centering
\begin{subfigure}{.48\textwidth}
  \centering
  \includegraphics[width=1\columnwidth]{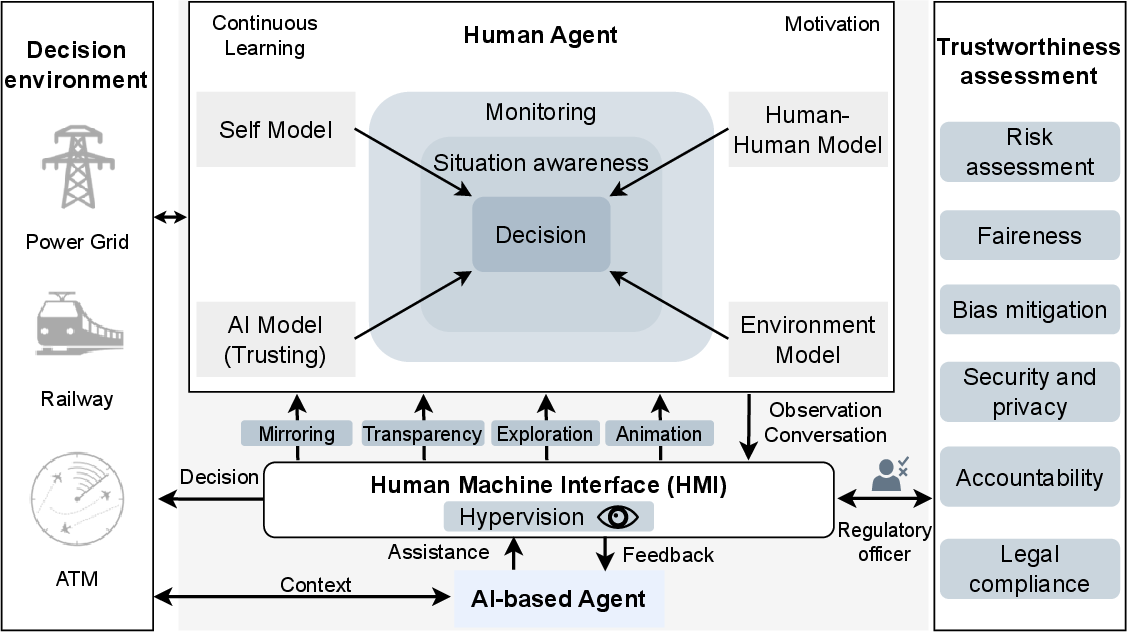}
  \caption{Human agent decision-making}
  \label{fig:human_dm}
\end{subfigure}
\hfil
\begin{subfigure}{.493\textwidth}
  \centering
  \includegraphics[width=1\columnwidth]{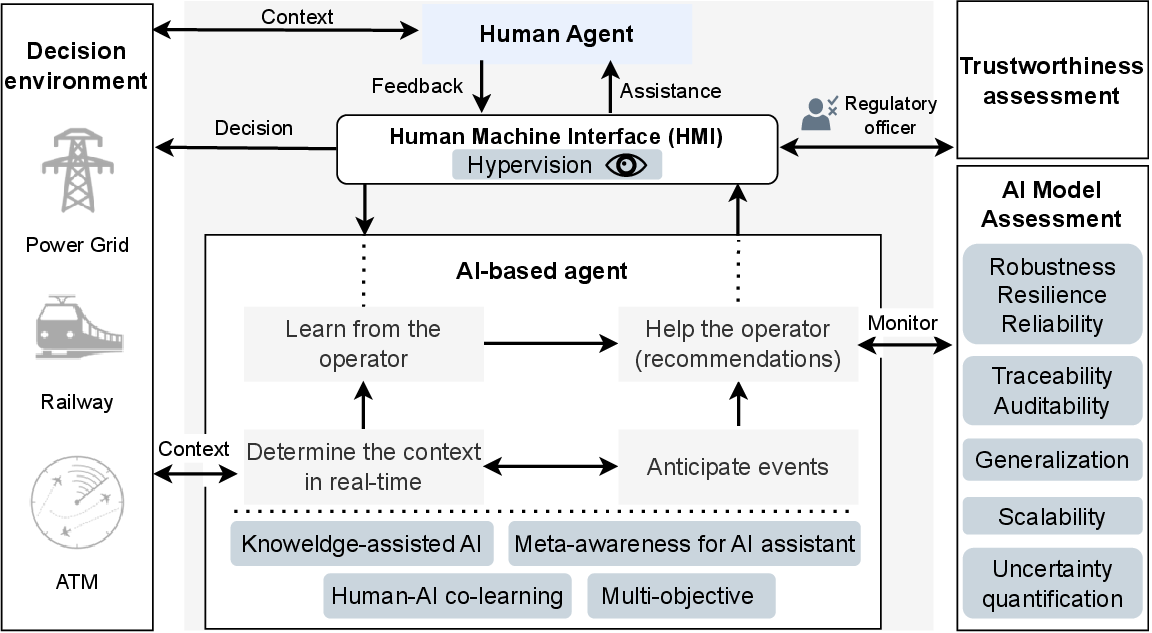}
  \caption{AI supported decision-making}
  \label{fig:ai_dm}
\end{subfigure}
\caption{Decision-making process (logical view) in critical network infrastructure operations presented from human (left) and AI (right) agents perspectives. The AI-based agent increases the human agent's cognitive processes through mirroring, transparency, exploration, and animation. Its performance is also monitored through various properties (e.g., robustness, generalization, scalability, etc.). The human agent, in turn, provides feedback to the AI-based agent, enabling its continuous improvement. The final decision is verified by a regulatory agent through trustworthy KPIs.}
\label{fig:decision-making}
\end{figure*}

\subsubsection{Human decision-making}
\label{sec:human_dm}
As illustrated in Figure \ref{fig:human_dm}, monitoring the operational process through AI support is essential for human to understand the situation and to develop adequate situation awareness \cite{endsley2023supporting}, which further increases the human cognitive process \cite{miller2018macrocognition}. The overarching goal is to continuously improve the human agent's knowledge required for decision-making, which is represented as mental models of the \textit{environment} (system and control knowledge), the \textit{AI} (knowledge about capabilities and limitations of AI-based assistant), the \textit{self} (awareness of their own strengths and weaknesses), and the \textit{human-human} collaboration (interrelations between decisions taken by individuals). The AI could support human learning by enhancing these domains of knowledge. 

The AI support is enabled through the Human Machine Interface (HMI) in four different ways: \textit{transparency} (provides interpretable and explainable assistance), \textit{exploration} (enables the humans to explore/learn a subject matter), \textit{animation} (requires the human to reflect or contribute), and \textit{mirroring} (AI mirrors individualized patterns in human behavior to make the human aware of own biases and variabilities in decision-making).  
It is crucial that the collaboration between humans and AI is deliberately designed and \textit{continuously} improved in such a way that it supports human learning processes and promotes human \textit{motivation} rather than overwhelming them. 


\subsubsection{AI-based decision-making}
\label{sec:ai_dm}
As can be seen in Figure \ref{fig:ai_dm}, the AI agent---
leveraging various learning paradigms such as knowledge-assisted AI, co-learning, multi-objective reinforcement learning, and meta-awareness for AI assistant---supports the human operator by determining the decision context within its complex underlying space in near real-time. Additionally, it enhances situation awareness by anticipating future events and providing recommendations. To ensure adaptability and continuous learning, the AI assistant must incorporate human feedback into its learning strategy and decision-making processes.

AI-based decision-making in safety-critical systems must also adhere to strict domain and regulatory standards to ensure safe operation and compliance.
These key requirements ensure the AI system can intervene immediately and effectively to maintain the system's normal operational state, even in the face of unexpected critical situations. Among these, \textit{generalization}~\cite{nichol2018gotta} and \textit{scalability}~\cite{ulanov2017modeling} are essential to adapt to varying conditions and handle diverse operational scenarios.
Furthermore, to ensure consistent performance across different circumstances, \textit{robustness}~\cite{behzadan2017whatever} and \textit{reliability}~\cite{zissis2019r3}, as outlined in the trustworthiness vocabulary standard (ISO/IEC TS 5723:2022), are indispensable considerations. Finally, the AI system must be designed to prevent, respond to, and recover from adversarial attacks, emphasizing the importance of \textit{resilience}.
Integrating these concepts during training and testing is crucial for the safe, reliable, and lawful operation of critical systems.

In human-AI decision-making, AI-based systems can augment human judgment with data-driven recommendations, enhancing efficiency and reducing bias. AI agents could also learn from human feedback and improve their recommendations. 
However, this synergy introduces new challenges in AI-based decision-making, requiring the AI to include additional characteristics: a) \textit{traceability} and \textit{auditability} to ensure the alignment between requirements and the product developed, and with the desired objectives and standards correspondingly; b) \textit{uncertainty quantification} to characterize uncertainties in AI models and real-world data, enabling humans to be aware of the limitations of the AI system; and, c) provide \textit{transparency} when interfacing with the human operator to understand the long-term impact of a certain decision \cite{gunning2019darpa}.



\subsubsection{Human-AI interaction}
\label{sec:interacion}
As a result of the logical analysis in the system engineering phase, HMI of the conceptual framework (see Figure \ref{fig:decision-making}) incorporates 
three distinct forms of interaction between human and AI-based agents, each representing a different level of collaboration and autonomy as described below: 

\begin{itemize}
    \item \textit{AI-assisted full human control}: AI offers high levels of automation in information acquisition, information integration, and possibly decision selection. Action implementation is fully allocated to the human operator. A practical example is where AI directs humans' attention to important system information, integrates it in intuitive and human-friendly ways, and offers (a set of) directions where good decisions should be made;
    \begin{figure}[H]
        \centering        
        \hspace{0.5cm}\includegraphics[width=0.7\linewidth,]{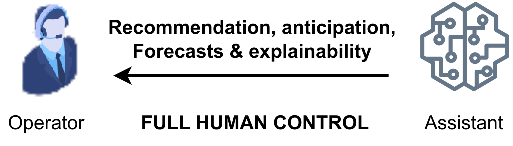}
        \label{fig:interaction_human}
    \end{figure}
    \item \textit{Joint human-AI decision-making}: AI and the human operator can both independently and autonomously observe information, make decisions, and undertake actions. In this configuration, bi-directional human-AI communication is required to ensure that both agents are aware of who is doing what, when, and how. A practical example is where the AI and human operator are working in parallel on completing a control task and, by observing each other’s behavior, can learn from each other. For co-learning, it may be necessary to consider lower levels of automation at the action implementation stage, where the AI provides specific advisories that the human can accept, reject, or modify. Furthermore, AI can go beyond the provision of recommendations and thereby explicitly support human cognitive processes of decision-making, learning, and motivation. This for instance by supporting humans in exploring their environment so that they learn to recognize weak signals for emerging problems and corresponding leverage points;
    \begin{figure}[H]
        \centering
        \hspace{0.5cm}
        \includegraphics[width=0.7\linewidth]{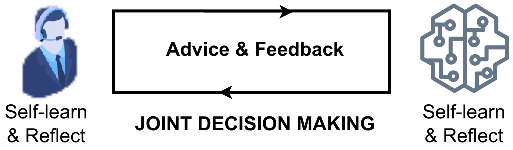}
        \label{fig:interaction_joint}
    \end{figure}
    \item \textit{Full AI-based control}: AI provides fully autonomous primitives that represent simple, intuitively understandable functionalities for human operators \cite{wahde2021five}. The human operator orchestrates these autonomous primitives to accomplish specific tasks. Ideally, operators do not need to intervene in these black-box primitives. However, in the event of system faults, the human-AI system must be able to fall back to lower levels of automation, allowing human intervention with the finest granularity of control.
    \begin{figure}[H]
        \centering
        \hspace{0.5cm}
        \includegraphics[width=0.7\linewidth]{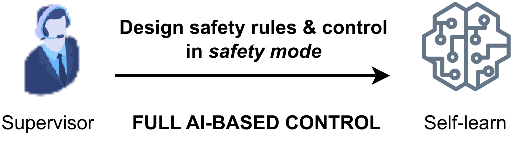}
        \label{fig:interaction_ai}
    \end{figure}
\end{itemize}

It is important to note that choosing the right levels and stages of automation is warranted by operational contexts, situational demands, and capabilities (and limitations) of human and automated agents. As such, a ``one-size-fits-all'' distribution of functions and tasks does not exist and will need to be re-considered per application domain and/or operational scenarios. Noteworthily, regulatory demands for critical systems include specific objectives on the levels of human autonomy and oversight. The effectiveness of each interaction mode can be assessed using both quantitative and qualitative KPIs, tailored to the system’s criticality. Key indicators include AI acceptability, trust, user experience, sociotechnical decision quality, and AI-human task allocation balance.\\ 

\noindent\textit{Hypervision}\quad  As infrastructure dynamics become more complex---such as with the energy transition in electric transmission systems---traditional supervision tools struggle to handle the growing volume of information. Managing multiple screens and applications places a cognitive burden on users, requiring them to manually prioritize and link disparate data before making decisions. The HMI module of the proposed framework integrates hypervision (see Figure \ref{fig:decision-making}), which synthesizes key information and centralizes real-time business events into a unified interface. This enhances decision-making and task prioritization by shifting the focus from alarm monitoring to efficient task execution \cite{amokrane2024framework, marot2022perspectives}.


\subsection{Philosophical Foundations of Trustworthy AI}
The collaborative process between the human operator and AI-based agent (see Figure \ref{fig:decision-making}) raises a discussion on the philosophical foundations of Trustworthy AI (TAI). Debates persist regarding the feasibility and conceptual legitimacy of attributing trustworthiness to AI \cite{nickel2010can}, given its lack of intrinsic motivations and moral obligations. These criticisms are addressed by reframing TAI’s ethical dimensions as compliance with specific ethical requirements rather than interpersonal trust analogies \cite{ryan2020ai}. This discussion is important not only for philosophical reasons but to provide the foundations of a systematic approach to risk assessment in AI-based systems, especially those used in safety-critical domains.


To address AI-related risks and uncertainties, the proposed framework incorporates a process for deriving ethically-aware functional and non-functional requirements and corresponding metrics and KPIs. This process adapts the European Commission’s ALTAI tool \cite{fedele2024altai} to be used for \textit{ex-ante} analysis of the AI system at different stages of the development cycle. It is complemented by a risk management approach that adapts the multi-component risk analysis commonly used in natural disaster management \cite{center2015sendai} to AI-related risks and technological contexts~\cite{zanotti2024ai}. This approach enables targeted interventions to mitigate risks through hazard reduction, exposure limitation, and decreasing vulnerability. The ALTAI framework, adapted for safety-critical applications, provides proactive guidelines across dimensions such as human oversight, technical robustness, societal well-being, and accountability. By reinterpreting ALTAI's self-assessment questions into actionable design-phase requirements, this framework fosters active responsibility, ensuring systems are technically sound and ethically aligned, minimizing risks like over-reliance and de-skilling in human-AI interactions.


\section{Case Study: Power Grid Management}
\label{sec:instantiation}

To demonstrate the effectiveness and domain-agnostic nature of the proposed framework, we apply it to the existing GridOptions framework \cite{Viebahn24}, which is currently used for decision support in a TSO control room. 



\begin{figure}
    \centering
    \includegraphics[width=1\columnwidth]{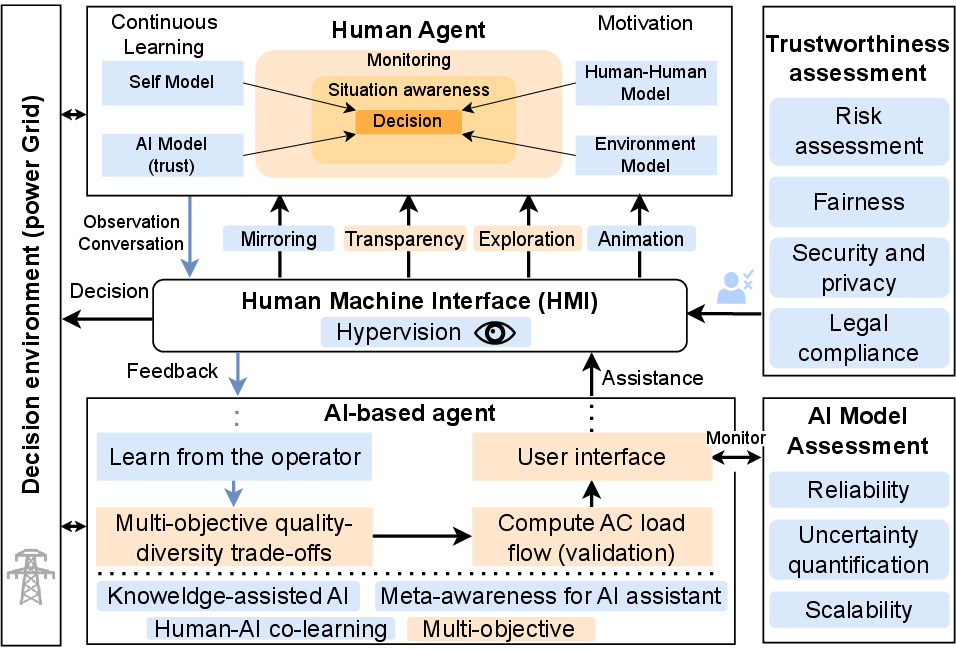}
    \caption{Conceptual framework instantiation for GridOptions. The original functionalities and flow of GridOptions frameworks are shown in orange color. The required enhancements through proposed conceptual framework are shown in blue}
    \label{fig:grid_options}
\end{figure}

\paragraph{System design}
At this stage, we analyze the functionalities of GridOptions framework, identify its limitations, and provide insights into how it can be enhanced through the proposed framework architecture, emphasizing the need for stronger human-AI interaction. For that purpose, Figure \ref{fig:grid_options} projects the GridOptions tool on the proposed conceptual framework by indicating both functionalities already present (orange) and functionalities not yet available (blue) in the tool.

\paragraph{Context and decision environment}

Line congestion poses a significant risk to power grid operation, as it can trigger cascading failures that may ultimately result in a large-scale blackout. The TSO is tasked with managing congestion on the transmission grid, which is characterized by extensive action and observation spaces due to the system's size. This management process involves sequential decisions across various time horizons, uncertainty (e.g., from weather-dependent generation sources such as renewables or measurement errors), behavioral diversity, and the pursuit of multiple objectives~\cite{Viebahn22}. Control centers provide groups of human operators with the necessary working and decision-making environment to remotely monitor the system and properly operate it in real time \cite{marot2022perspectives}. However, deploying AI-based decision support tools in TSO control rooms is still in its infancy.

The GridOptions tool represents one of the first AI-based decision support tools that is deployed in a control room \cite{Viebahn24}. It recommends to operators remedial actions to prevent congestion in the intraday timeframe (i.e., within a 24-hour forecast horizon). However, the scope of the first version of the GridOptions tool has been limited to enable fast development and release cycles. Consequently, the decision-making context can be enlarged in several ways. For example, increasing the geographical scope and adding other kinds of remedial actions can still vastly increase the observation and action spaces. This directly relates to the issues of scalability and generalization  (sec.~\ref{sec:ai_dm}). Moreover, the GridOptions tool currently represents a standalone tool with a simple user interface. In the future, the tool needs to be integrated with several other decision support functionalities in the control room, and it needs to integrate several different time horizons (e.g., from operation planning to real-time). This directly relates to the issues of information overload and hypervision (see sec.~\ref{sec:interacion}).

\paragraph{Decision-making}

Regarding the human agent, the question is how far AI-based decision support enhances crucial human cognitive processes (decision-making, continuous learning, appropriate trust, intrinsic motivation) via AI model capabilities (transparency, exploration, animation, and mirroring) (sec.~\ref{sec:human_dm}). Currently, the GridOptions tool mainly aims to support human cognitive decision-making processes. On top of monitoring and situation awareness support tools, which are commonly available in power system control rooms, the GridOptions tool offers operators a set of different plans to mitigate congestion. 
In contrast, the cognitive processes of learning, trust, and intrinsic motivation are not directly addressed in the current version of the GridOptions tool. Hence, there are still several directions in which the GridOptions tool could be improved. For example, enabling the operator to explore their own assumptions and plans (i.e., the operator can insert self-generated remedial actions and compare those with recommended remedial actions) and causal relations in the environment can support both continuous learning and intrinsic motivation. Moreover, animation capabilities in the form of alerts (e.g., when previously unseen remedial actions are recommended), questions (e.g., whether recommended remedial actions are consistent with the operators' background knowledge), and mirroring (reflecting the operator's decision-making style), can support learning and motivation.

Regarding AI-based decision-making (sec.~\ref{sec:ai_dm}), the GridOptions tool employs multi-objective quality-diversity optimization, which enables the generation of a set of plans that satisfy different trade-offs between various objectives and are behaviourally diverse. The considered objectives are related to both infrastructure security constraints and the complexity of the plans.
Presenting the advantages and disadvantages of the different plans in an easily comparable fashion fosters AI explainability.

However, a broad range of improvements are still possible or even necessary. For example, uncertainty quantification, robustness, and reliability quantification are still needed both for rigorous AI model assessment and for enabling adequate human trust.
Moreover, learning paradigms like co-learning and meta-awareness still need to be developed to enable smooth and effective human-AI collaboration.

Finally, regarding the human-AI interaction, the GridOptions tool only offers the AI-assisted full human control mode of interaction (see sec.~\ref{sec:interacion}). Before the introduction of the GridOptions tool, operators were fully in control of action suggestions and action implementation. The introduction of AI-based decision support tooling needs to proceed incrementally to enable learning and trust and thus assure motivation on the human agent side (see sec.~\ref{sec:human_dm}). Only after the current version of the GridOptions tool has sufficiently demonstrated reliability, robustness, and explainability (see sec.~\ref{sec:ai_dm}) more advanced modes of interaction can be explored. For that to happen, more research is needed, such as how co-learning and smooth feedback cycles can be made possible.

\paragraph{Trustworthiness assessment}


For GridOptions, many of these advancements are still in the early stages and require further research and development. We are actively monitoring the evolving EU AI Act and a fundamental first step in this process is conducting a comprehensive risk assessment to evaluate potential hazards, vulnerabilities, and compliance requirements. This will be essential to secure long-term operational integration while increasing trust in AI-based systems.

\section{Conclusions}
\label{sec:conclusions}
This paper presents a conceptual and technology-agnostic framework developed through a collaborative effort between academia and industry, including operators from three distinct critical infrastructures. Designed to integrate AI into decision-making processes, the framework ensures a balanced approach between automation and human oversight. It fosters effective human-AI collaboration through an iterative joint-learning process, enabling operators to continuously refine AI behavior and enhance decision-making. Furthermore, it supports real-time operations by using integrated data and predictive analytics, enabling both proactive and corrective interventions across different levels of automation.

Beyond the technical aspects, this work advocates for focusing not only on individual AI models but also on jointly designing and optimizing the broader socio-technical system in which AI operates. Successful adoption of emerging technologies depends on an interdisciplinary approach, shifting from isolated decision-making to collaborative human-AI interaction. Insights from disciplines such as philosophy, ethics, and cognitive engineering help ensure that AI-based systems are developed and deployed with a deep understanding of their operational and societal contexts.



\bibliographystyle{IEEEtran}
\bibliography{references}

\end{document}